\documentclass[final,a4paper]{aastex702}

\usepackage{amsmath}

\usepackage{etoolbox}

\newcommand{\titleherofigure}{%
  \par\vspace{10pt}%
  \begin{center}
    \includegraphics[width=0.96\textwidth]{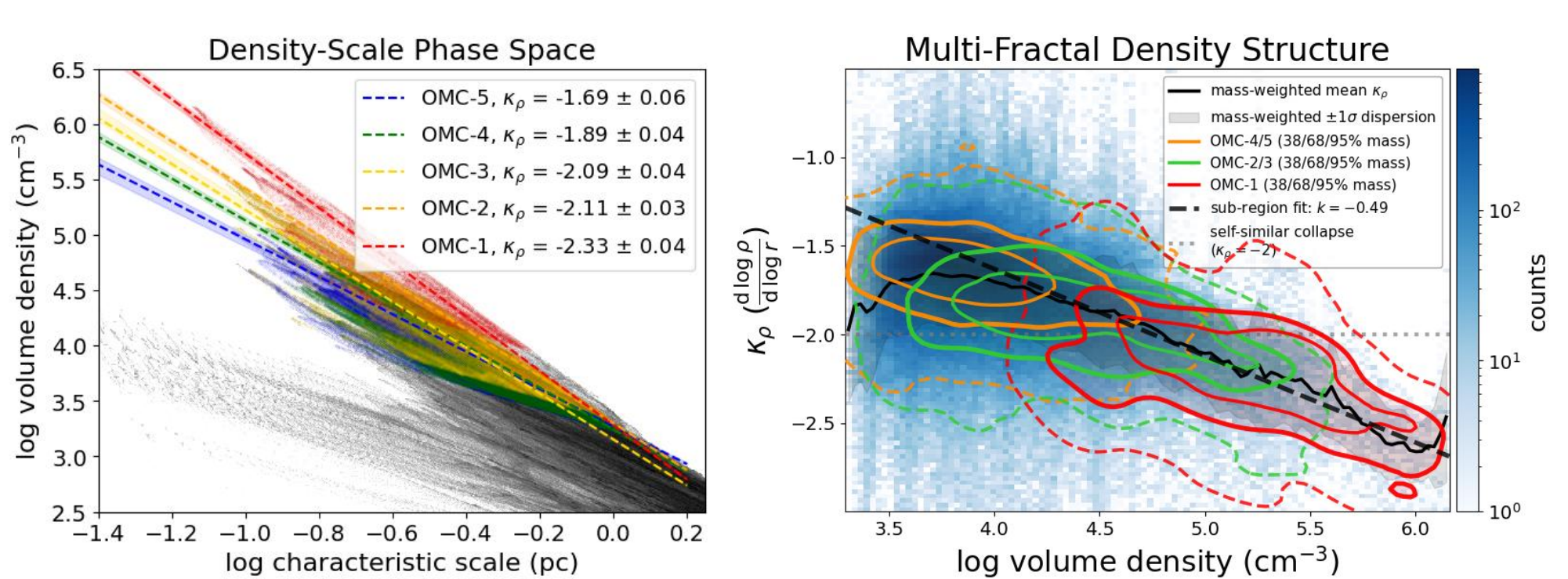}%
  \end{center}
  \vspace{4pt}%
  \noindent\textit{Multi-fractal density structure of the Orion A Integral Shaped Filament.}Multi-fractal density structure of the Orion A Integral Shaped Filament. Left: Regional density–scale relations steepen systematically from OMC-4/5 to OMC-1. Right: The pixel-level distribution shows a continuous progression toward higher density and more negative density exponent $\kappa_\rho$.
  \par\vspace{12pt}%
}

\makeatletter
\patchcmd{\titleblock@produce}
  {\frontmatter@abstract@produce}
  {\titleherofigure\frontmatter@abstract@produce}
  {}{\PackageError{SVA}{Could not insert title hero figure}{}}
\makeatother

\begin{document}

\title{Gravity-driven Emergence of Multi-fractal Density Structure in the Orion A Integral Shaped Filament}
\shorttitle{Multi-fractal Density Structure in Orion A}
\shortauthors{Zhao et al.}

\correspondingauthor{Guang-Xing Li, Keping Qiu}

\author[orcid=0000-0003-0596-6608]{Mengke Zhao}
\affiliation{School of Astronomy and Space Science, Nanjing University, 163 Xianlin Avenue, Nanjing 210023, Jiangsu, People's Republic of China}
\affiliation{Key Laboratory of Modern Astronomy and Astrophysics (Nanjing University), Ministry of Education, Nanjing 210023, Jiangsu, People's Republic of China}
\email{mkzhao@nju.edu.cn}

\author[orcid=0000-0003-3144-1952]{Guang-Xing Li}
\affiliation{South-Western Institute for Astronomy Research, Yunnan University, Kunming 650091, People's Republic of China}
\email[show]{gxli@ynu.edu.cn}
\email[show]{ligx.ngc7293@gmail.com}

\author[orcid=0000-0002-5093-5088]{Keping Qiu}
\affiliation{School of Astronomy and Space Science, Nanjing University, 163 Xianlin Avenue, Nanjing 210023, Jiangsu, People's Republic of China}
\affiliation{Key Laboratory of Modern Astronomy and Astrophysics (Nanjing University), Ministry of Education, Nanjing 210023, Jiangsu, People's Republic of China}
\email[show]{kpqiu@nju.edu.cn}

\author[orcid=0009-0000-3140-8955]{Guangya Zeng}
\affiliation{South-Western Institute for Astronomy Research, Yunnan University, Kunming 650091, People's Republic of China}
\email{zengguangya@mail.ynu.edu.cn}

\begin{abstract}
Molecular clouds are often described as self-similar structures, although spatially averaged measures do not retain local variations in density scaling. We use the density exponent $\kappa_\rho$ ($\rho \propto r^{\kappa_\rho}$) to characterize the density structure of the Integral Shaped Filament (ISF) in Orion\,A. Applying the Multiscale Decomposition Reconstruction method to the Herschel column-density map, we find distinct density--scale relations across the connected filament. Their slopes steepen from $\kappa_\rho \approx -1.7$ to $-1.9$ in the quiescent OMC-4/5 regions, through $\approx -2.1$ in the star-forming OMC-2/3, to $\approx -2.3$ in OMC-1, which hosts massive star formation. The ISF therefore does not follow a single local density-scaling exponent but exhibits multi-fractal density scaling. The pixel-level distributions show the same progression toward higher volume density and more negative $\kappa_\rho$. Since $\kappa_\rho$ measures the concentration of gas toward smaller scales, we interpret this sequence as gravity-driven differential collapse: denser regions have shorter free-fall times and develop steeper density profiles. Longitudinal gas motions toward OMC-1 may limit the mass supply available for large-scale growth in the outer sub-regions and help maintain the observed range of local exponents. These results link local density scaling to gravitational concentration within a single connected filamentary system.
\end{abstract}

\keywords{\uat{Interstellar filaments}{842} --- \uat{Interstellar medium}{847} --- \uat{Molecular clouds}{1072} --- \uat{Star formation}{1569}}

\section{Introduction}

Molecular clouds are commonly described as self-similar, hierarchical structures \citep{1981MNRAS.194..809L,1987ARA&A..25...23S,2007ARA&A..45..565M}, although multifractal scaling has long been recognized in the cool ISM \citep{2001ApJ...551..712C,2018MNRAS.481..509E}. Statistical measures such as the density probability distribution function have been widely used to characterize cloud density structure \citep{2013ApJ...766L..17S,2019MNRAS.490.3061V}. Because these statistics average over position, however, they cannot by themselves distinguish regions with different local scaling exponents \citep{2022MNRAS.514L..16L}.

There are physical reasons to expect such variation. Gravitational collapse steepens the density profile, and the collapse rate depends on the local density ($t_{\rm ff} \propto \rho^{-1/2}$). Self-gravitating turbulence simulations show that gravity produces a high-density power-law tail in the density PDF and changes the density structure of the gas \citep{2011ApJ...727L..20K,2011MNRAS.416.1436B,2014ApJ...781...91G}. Denser regions therefore collapse on shorter timescales and develop steeper profiles than their less dense surroundings \citep{2021MNRAS.502.4963G}. The density exponent $\kappa_\rho$, defined through $\rho \propto r^{\kappa_\rho}$, quantifies this scaling. It is related to the mass--scale relation $M \propto r^{\kappa_\rho+3}$ \citep{1981MNRAS.194..809L} and traces the concentration of mass with scale. A physically useful reference value is $\kappa_\rho = -2$, corresponding to scale-free gravitational collapse in the approximately spherical/isotropic case \citep{1977ApJ...214..488S,2018MNRAS.477.4951L} and, for an approximately isothermal hierarchy, to a scale-independent Jeans ratio \citep{1902RSPTA.199....1J,2017MNRAS.465..667L}. For $\kappa_\rho < -2$, the enclosed Jeans ratio $M/M_J \propto r^{3+3\kappa_\rho/2}$ increases toward smaller scales, favoring super-Jeans fragmentation \citep{2025MNRAS.541.3869L,2026arXiv260717215M}. If $\kappa_\rho$ varies spatially within a cloud, the density structure is multi-fractal rather than mono-fractal \citep{2001ApJ...551..712C,2018MNRAS.481..509E}. Here we use ``multi-fractal density scaling'' in the phenomenological sense of spatially varying local density-scaling exponents; this is related to, but not identical with, a full multifractal-spectrum analysis.

Previous work has shown that density structure varies with evolutionary state within individual clouds. \citet{2015A&A...577L...6S} found that the N-PDF slope in Orion\,A correlates with the Class~0 protostar fraction, linking density structure to star-formation activity. \citet{2022MNRAS.514L..16L} introduced a Level-Set formalism to map $\kappa_\rho$ across the Perseus molecular cloud and found steeper profiles in denser sub-regions. Perseus, however, consists of loosely connected clumps with potentially different formation histories. A connected filament spanning a range of evolutionary states allows the density scaling to be compared within a common large-scale environment.

The Multiscale Decomposition Reconstruction method \citep[MDR,][]{2026ApJ...997..345Z} resolves the density--scale hierarchy from a column-density map. Its constrained diffusion algorithm \citep{2022ApJS..259...59L} decomposes nested column-density structures across spatial scales and assigns each pixel a characteristic scale $l_c$ and volume density $n_H = \Sigma / l_c$. Throughout the main text, $\Sigma$ denotes the H$_2$ column number density in units of cm$^{-2}$.

The Integral Shaped Filament (ISF) in Orion\,A ($d \approx 400$\,pc; \citealt{2007A&A...474..515M}) contains five sub-regions (OMC-1 through OMC-5) within the same connected filamentary system \citep{2018A&A...619A.106G}. They span quiescent gas (OMC-4/5), active star formation (OMC-2/3; \citealt{2012AJ....144..192M}), and massive star formation (OMC-1; \citealt{2017A&A...602L...2H}). The global ISF density profile is approximately a power law, with $\Sigma(b) \propto b^{-5/8}$ and an equivalent radial profile $\rho(r) \propto r^{-13/8} \approx r^{-1.6}$ \citep{2016A&A...590A...2S}. Higher-resolution ArT\'eMiS observations resolve Plummer-like radial structure in the northern ISF \citep{2021A&A...651A..36S}. Recent work has extended filament-profile measurements to scale-dependent surface- and volume-density structure across nearby clouds, including Orion\,A \citep{2026A&A...710A.199Z}, and to the relation between projected surface-density and intrinsic volume-density profiles \citep{2026arXiv260412570M}. We use MDR to determine whether this global relation represents a common local density scaling across the filament. Longitudinal gas motions toward OMC-1 \citep{2017A&A...602L...2H} also place the density structure in the context of mass redistribution along the ISF.

\section{Data and Method}

We use the H$_2$ column density map of Orion\,A from the Herschel Gould Belt Survey\footnote{\url{http://www.herschel.fr/cea/gouldbelt/en/Phocea/Vie_des_labos/Ast/ast_visu.php?id_ast=66}} \citep{2010A&A...518L...2P,2010A&A...518L...3G,2013ApJ...763...55R,2013ApJ...777L..33P}, derived via SED fitting at wavelengths of 70, 160, 250, 350, and 500\,$\mu$m, at a resolution of $36''$ ($\sim 0.07$\,pc at 400\,pc). Throughout the main text, $\Sigma$ denotes the H$_2$ column number density in units of cm$^{-2}$. The MDR method \citep{2026ApJ...997..345Z} uses constrained diffusion \citep{2022ApJS..259...59L} to decompose the column-density field into components $N_i(x,y)$ associated with spatial scales $r_i$, from which the characteristic scale $l_c$ and volume density $n_H$ are reconstructed at each pixel:
\begin{equation}
\Sigma(x,y)
\;\xrightarrow{\;\mathrm{MDR}\;}\;
\{N_i(x,y),\,r_i\}
\;\longrightarrow\;
l_c(x,y),
\qquad
n_H(x,y)=\frac{\Sigma(x,y)}{l_c(x,y)}.
\label{eq:mdr}
\end{equation}
Here $l_c$ provides a calibrated estimate of the effective line-of-sight thickness. MDR has been validated against both filamentary and complex MHD cloud structures across multiple projection directions, with reconstructed volume densities centered on the true values and dispersions of 0.21--0.28\,dex \citep{2026ApJ...997..345Z}. Implementation details are given in Appendix~\ref{app:mdr_details}.

At the sub-region level, we fit $\log n_H$ versus $\log l_c$ using principal-component analysis (equivalent to total least squares), with the slope of the first principal component giving $\kappa_\rho$ for each sub-region. Systematic uncertainties are quantified via sensitivity analysis (Appendix~\ref{app:mdr_details}). At the pixel level, the Adjacent Correlation Analysis \citep[ACA,][]{2025arXiv250605759L,2025arXiv250605758L} maps $\kappa_\rho$ across the cloud. At each position, ACA evaluates the spatial gradients of $\log n_H$ and $\log l_c$, combines the gradient vectors from different spatial directions via a pseudo-Stokes formalism to obtain a local correlation angle $\theta$, and returns the pixel-level density exponent $\kappa_\rho$, defined through
\begin{align}
\rho &\propto r^{\,\kappa_\rho}, \label{eq:kappa_def}\\[4pt]
\kappa_\rho &= \frac{d\log\rho}{d\log r}
             = \frac{d\log n_H}{d\log l_c}
             = \tan\theta,
\label{eq:kappa_aca}
\end{align}
where $\theta$ is the ACA correlation angle.
The pixel-level distributions are weighted by column density to suppress diffuse regions where local gradients are noisier. The full ACA equations are given in Appendix~\ref{app:aca_equations}. The identification of $d\log n_H / d\log l_c$ with the three-dimensional radial density exponent $d\log\rho / d\log r$ is calibrated against synthetic observations in Appendix~\ref{app:mdr_aca_validation}, where structures with different known input exponents remain separated after projection, MDR reconstruction, and ACA measurement.

The five sub-regions are delineated using spatial masks consistent with the ISF structure identified in previous studies \citep{2017A&A...602L...2H,2018A&A...619A.106G}. The sub-region masks and density--scale fits adopt $\Sigma > 10^{21.8}\,{\rm cm}^{-2}$, which traces the coherent ISF ridge; the pixel-level $\kappa_\rho$--$n_H$ distributions adopt $\Sigma > 3\times 10^{21}\,{\rm cm}^{-2}$ to sample a wider density range, with column-density weighting to reduce the contribution of diffuse pixels. As an MDR-independent measure of regional mass concentration, we compute
\begin{align}
\langle\rho\rangle &= \frac{M}{(L/2)^3}, \label{eq:rhobar}
\end{align}
where $M$ is the total gas mass above $\Sigma > 10^{22}\,{\rm cm}^{-2}$ from integrating the Herschel map (independent of MDR) and $L$ is the filamentary skeleton length. This quantity is not a literal three-dimensional mean density; it is used as a characteristic measure of mass concentration for the five sub-regions. The numerical values are listed in Table~\ref{tab:subregion}.

\begin{table}
\centering
\caption{Properties of the ISF sub-regions. $M$ is the gas mass above $\Sigma > 10^{22}$\,cm$^{-2}$, $L$ the filamentary skeleton length, $\langle\rho\rangle = M/(L/2)^3$ the characteristic density proxy, and $\kappa_\rho$ the fitted density exponent with its sensitivity-analysis uncertainty.}
\label{tab:subregion}
\begin{tabular}{lcccc}
\hline\hline
Region & $M$ ($M_\odot$) & $L$ (pc) & $\langle\rho\rangle$ $(M_\odot\,\mathrm{pc}^{-3})$ & $\kappa_\rho$ \\
\hline
OMC-5 & 5444 & 8.620 & 68   & $-1.69 \pm 0.06$ \\
OMC-4 & 2404 & 4.988 & 155  & $-1.89 \pm 0.04$ \\
OMC-3 & 1999 & 2.891 & 662  & $-2.09 \pm 0.04$ \\
OMC-2 & 2048 & 2.848 & 709  & $-2.11 \pm 0.03$ \\
OMC-1 & 4650 & 3.033 & 1333 & $-2.33 \pm 0.04$ \\
\hline
\end{tabular}
\end{table}

\begin{figure*}
\centering
\includegraphics[width=0.9\linewidth]{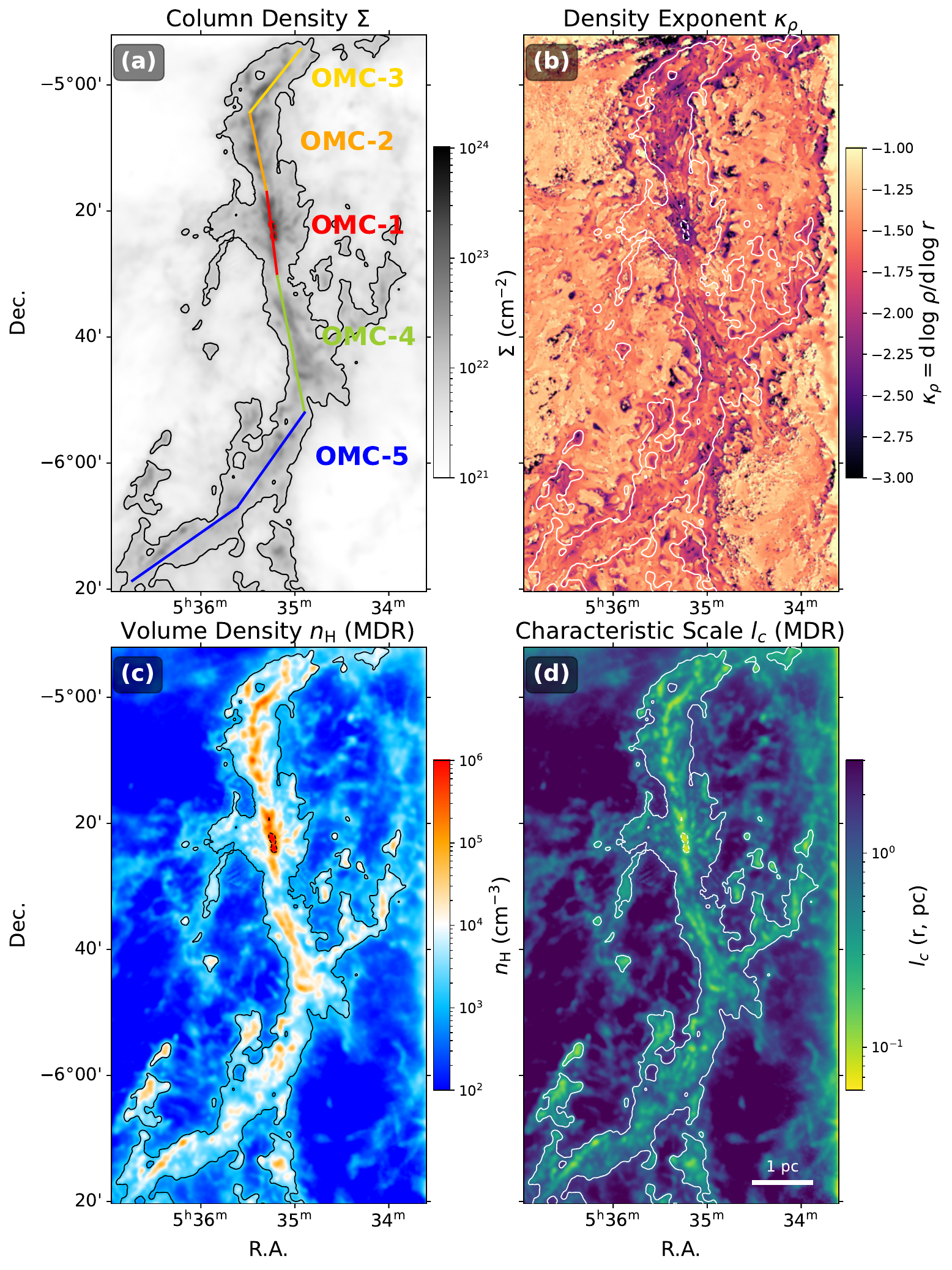}
\caption{Density structure of the ISF in Orion A. 
(a) H$_2$ column density from the Herschel Gould Belt Survey. Colored lines mark the skeletons used to determine the characteristic length $L$ and define the regional mass measurements. The contour outlines $\Sigma = 10^{21.8}$\,cm$^{-2}$, the minimum threshold for MDR analysis. 
(b) Spatially resolved density exponent $\kappa_\rho = \mathrm{d}\log\rho/\mathrm{d}\log r$, computed from MDR-derived $(n_H, l_c)$ via ACA. Pixels with $\Sigma < 10^{21.8}$\,cm$^{-2}$ are masked. The dashed line in the colorbar marks $\kappa_\rho = -2$. Yellow (dark) colors indicate shallow (steep) density profiles.
(c) MDR-predicted volume density $n_H = \Sigma/l_c$, achieving $\pm 0.25$\,dex accuracy \citep{2026ApJ...997..345Z}.
(d) MDR-derived characteristic scale $l_c$. The white bar indicates 1\,pc at 400\,pc. White contours in (c)--(d) mark $\Sigma = 10^{21.8}$\,cm$^{-2}$.
\label{fig1}}
\end{figure*}

\begin{figure*}
\centering
\includegraphics[width=\linewidth]{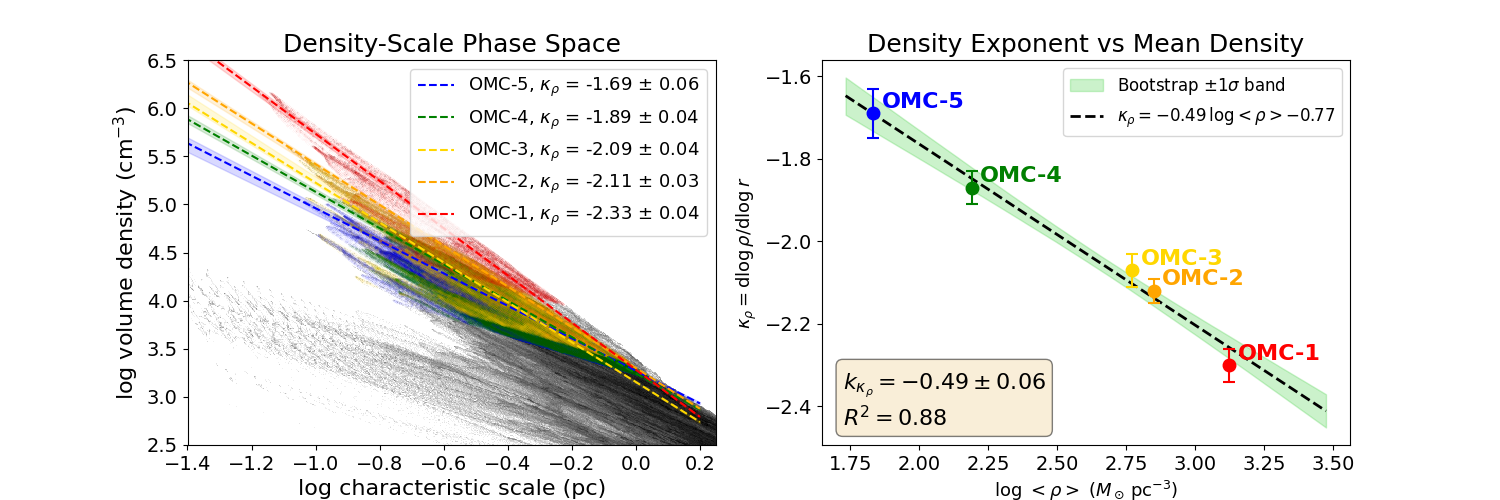}
\caption{Left: Density--scale phase space for each ISF sub-region. Colored points show pixels above the adopted column-density threshold, and dashed lines show the fitted density--scale relations. The fitted $\kappa_\rho$ values and their sensitivity-analysis uncertainties are listed in the legend.
Right: Density exponent versus the characteristic density proxy $\langle\rho\rangle = M/(L/2)^3$, calculated independently of MDR. The dashed line is a weighted linear fit and the green band is its bootstrap $1\sigma$ confidence interval. The fitted slope is $k_{\kappa_\rho} = -0.49 \pm 0.06$, with $R^2 = 0.88$.
\label{fig2}}
\end{figure*}

\begin{figure}
\centering
\includegraphics[width=0.7\linewidth]{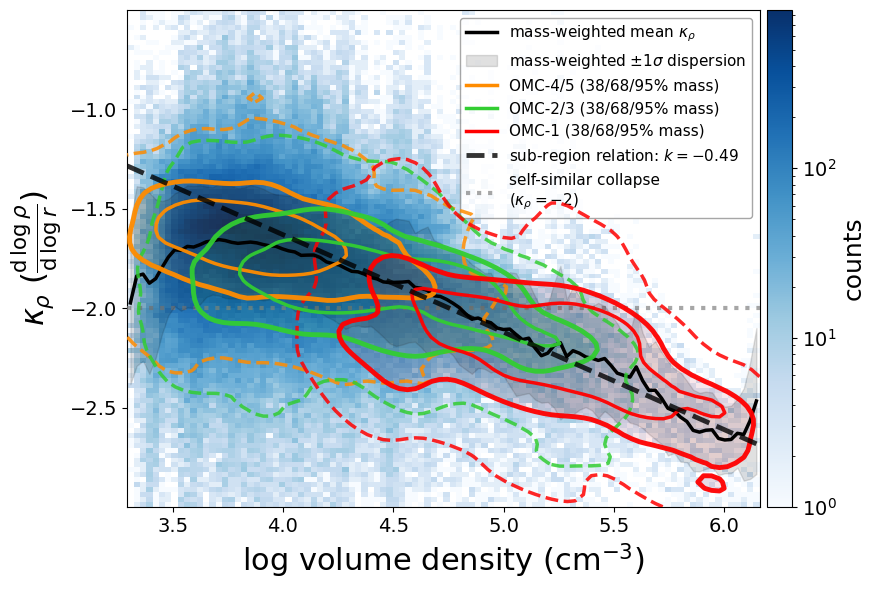}
\caption{Multi-fractal density structure of the ISF. Density exponent $\kappa_\rho$ versus volume density $n_{\rm H}$ for all selected ISF pixels ($\Sigma > 3\times 10^{21}$\,cm$^{-2}$). The blue two-dimensional histogram shows the unweighted pixel counts. The black curve and gray band show the column-density-weighted mean $\kappa_\rho$ and its weighted $1\sigma$ dispersion in each density bin. For equal-area pixels, column-density weighting is equivalent to gas-mass weighting. 
The colored KDE contours enclose 38\%, 68\%, and 95\% of the weighted mass distributions of OMC-4/5, OMC-2/3, and OMC-1.
The groups occupy different regions of this plane, tracing the spatial variation in density scaling across the ISF. The dotted horizontal line marks $\kappa_\rho = -2$. The dashed diagonal line shows the slope $k_{\kappa_\rho} = -0.49$ measured from the sub-region relation in Figure~\ref{fig2}; its vertical offset is chosen only for visual comparison because the two figures use different density variables.
\label{fig3}}
\end{figure}

\section{Results and Discussion}

\subsection{Multi-fractal Density Structure}

The five ISF sub-regions follow different density--scale relations (Fig.~\ref{fig2}, left panel). The fitted exponent changes from $\kappa_\rho = -1.69$ in OMC-5 to $-1.89$ in OMC-4, $-2.09$ and $-2.11$ in OMC-3 and OMC-2, and $-2.33$ in OMC-1 (Fig.~\ref{fig1}; Table~\ref{tab:subregion}). OMC-4/5 therefore have relatively shallow profiles with $\kappa_\rho > -2$, OMC-2/3 lie just below the $-2$ reference, and OMC-1 has the steepest density profile.

The distinct slopes rule out a single local density-scaling exponent for the ISF. The global profile $\rho \propto r^{-1.6}$ \citep{2016A&A...590A...2S} is therefore an average over regional variations in density scaling. In this sense, the connected filament exhibits multi-fractal density structure rather than a single self-similar density hierarchy.

\subsection{Density Exponent Steepens with Density}
\label{sec:kappa_density}

The same sequence appears in the MDR-independent characteristic-density proxy: denser sub-regions have more negative $\kappa_\rho$ (Fig.~\ref{fig2}, right panel). A weighted linear fit gives
\begin{equation}
\kappa_\rho = (-0.49 \pm 0.06)\,\log\!\left(\frac{\langle\rho\rangle}{M_\odot\,\mathrm{pc}^{-3}}\right) -0.77,
\label{eq:kappa_rho_fit}
\end{equation}
with $R^2 = 0.88$. The relation traces a progression from lower-density regions with shallow profiles to denser regions with stronger concentration toward small scales. Within the dense ISF hierarchy, this is the progression expected from gravitational collapse: denser regions have shorter free-fall times ($t_{\rm ff} \propto \rho^{-1/2}$) and develop steeper density profiles \citep{2021MNRAS.502.4963G}. The $\kappa_\rho = -2$ reference corresponds to scale-free collapse in the approximately spherical/isotropic case \citep{1977ApJ...214..488S,2018MNRAS.477.4951L}; for $\kappa_\rho < -2$, super-Jeans fragmentation is favored \citep{2025MNRAS.541.3869L,2026arXiv260717215M}, consistent with the massive star formation observed in OMC-1. The trend is broadly consistent with the density--structure relation found in Perseus \citep{2022MNRAS.514L..16L}.

This regional ordering persists at the pixel level (Fig.~\ref{fig3}): the column-density-weighted distributions of OMC-4/5, OMC-2/3, and OMC-1 shift progressively toward higher $n_{\rm H}$ and more negative $\kappa_\rho$, and the mass-weighted mean relation steepens continuously with increasing density. The high-probability locus of OMC-1 lies predominantly below $\kappa_\rho = -2$. The continuous distribution of local density-scaling exponents within one connected cloud is the sense in which we refer to the structure as multi-fractal \citep{2001ApJ...551..712C,2018MNRAS.481..509E}.

\subsection{Mass Redistribution along the ISF}

Observed longitudinal gas motions toward OMC-1 across OMC-2/3/4 \citep{2017A&A...602L...2H} provide a channel for mass redistribution along the filament. OMC-5 follows the same density--scale sequence, although its participation in the measured flow is less well constrained. Gas transported toward OMC-1 can sustain mass accumulation and profile steepening there while reducing the material available for large-scale growth in the outer filament. Longitudinal transport can therefore limit the development of the outer sub-regions toward an OMC-1-like hub without suppressing local collapse, fragmentation, or star formation, which remain active in OMC-2/3. Differential mass supply may thus help maintain the observed range of local density exponents. Quantifying the regional mass budget requires dedicated kinematic modeling.

\section{Conclusion}

The ISF in Orion\,A contains distinct density--scale relations, with $\kappa_\rho$ ranging from $\approx -1.7$ in OMC-4/5 to $\approx -2.3$ in OMC-1. The global exponent ($\approx -1.6$) inferred from spatially averaged measurements therefore does not represent a common local scaling across the filament. Instead, the ISF contains a continuous distribution of local density exponents and exhibits multi-fractal density scaling.

An MDR-independent measure of mass concentration gives the same regional sequence: denser sub-regions have steeper density profiles. We interpret this sequence as gravity-driven differential collapse, in which shorter free-fall times at higher density lead to stronger concentration toward small scales. Longitudinal mass transport toward OMC-1 may help maintain the range of exponents by reducing the material available for large-scale growth in the outer filament.

The same $\kappa_\rho$--density trend appears in Perseus \citep{2022MNRAS.514L..16L}, suggesting that spatially varying density exponents may be a general feature of the molecular ISM \citep{2001ApJ...551..712C,2018MNRAS.481..509E}.


\bibliography{reference}
\bibliographystyle{aasjournalv7.1}

\appendix

\section{MDR Implementation Details}
\label{app:mdr_details}

The MDR method \citep{2026ApJ...997..345Z} is the two-dimensional reconstruction mode of the Volume Density Mapper code (VDM; \citealt{2025arXiv250917369L}). The constrained diffusion algorithm \citep{2022ApJS..259...59L} decomposes the column density into components $N_i(x,y)$ at specific scales $r_i = 2^0, 2^1, \ldots, 2^{n_{\max}}$ pixels. The characteristic scale is the intensity-weighted average in log space:
\begin{equation}
\log_2 l_c = \frac{\sum_i N_i(x,y) \cdot \log_2 r_i}
                   {\sum_i N_i(x,y)}.
\tag{A1}
\end{equation}

The conversion from column density to volume density assumes that the MDR characteristic scale provides a calibrated estimate of the effective line-of-sight depth ($l_{\rm depth} \approx l_c$), so that $n_H = \Sigma / l_c$. Because the line-of-sight thickness of a molecular cloud is not directly observable \citep{2015ApJ...811...71Q}, the identification $l_{\rm depth} \simeq l_c$ requires empirical calibration. $l_c$ is a correlated scale validated to match the FWHM of the line-of-sight density profile in simulations \citep{2026ApJ...997..345Z}.

The constrained diffusion algorithm preserves the observed plane-of-sky morphology and does not require spherical symmetry \citep{2022ApJS..259...59L}. For an elongated ridge, the stronger transverse curvature causes $l_c$ to track the local filament width, i.e., the width of the structure that dominates the column density at that position, while longitudinal morphology and nested substructures remain in the component maps. The identification of $l_c$ with an effective line-of-sight thickness has been calibrated across multiple projection directions \citep{2026ApJ...997..345Z}, with a typical density uncertainty of ${\sim}0.25$\,dex. The recovered density--scale slopes remain stable within this uncertainty across the morphologies and projections tested by \citet{2026ApJ...997..345Z}.

Three column-density thresholds are used for distinct purposes. The sub-region masks and density--scale fits adopt $\Sigma > 10^{21.8}\,{\rm cm}^{-2}$, which traces the coherent ISF ridge. The regional masses used to construct the characteristic density proxy are integrated above $\Sigma > 10^{22}\,{\rm cm}^{-2}$ within the same spatial masks, so that $\langle\rho\rangle$ emphasizes the dense material associated with each sub-region. The pixel-level $\kappa_\rho$--$n_H$ distributions adopt $\Sigma > 3\times 10^{21}\,{\rm cm}^{-2}$ to sample a wider density range; column-density weighting reduces the contribution of the additional low-column-density pixels.

Systematic uncertainties on the sub-region $\kappa_\rho$ values are quantified by a sensitivity analysis in which the column density threshold is varied by $\pm 0.15$\,dex and the spatial mask boundaries by $\pm 50$ pixels, yielding 25 slope estimates per sub-region; $\sigma(\kappa_\rho)$ is taken as the half-width of the 16th--84th percentile range of these estimates.

\section{ACA Pseudo-Stokes Equations}
\label{app:aca_equations}

The Adjacent Correlation Analysis \citep[ACA,][]{2025arXiv250605759L,2025arXiv250605758L} provides a pixel-level view of $\kappa_\rho$. At each position $\vec{x}$, the spatial gradients of two quantities $p_1 = \log n_H$ and $p_2 = \log l_c$ define a local correlation vector in the phase space:
\begin{equation}
\vec{G}_i(\vec{x}) = \left(\frac{\partial p_1}
{\partial x_i}, \frac{\partial p_2}{\partial x_i}\right).
\tag{B1}
\end{equation}
To combine these gradient vectors from different spatial directions, ACA treats them as spin-2 quantities and computes pseudo-Stokes parameters:
\begin{equation}
I = \sum_i (G_{1,i}^2 + G_{2,i}^2), \quad
Q = \sum_i (G_{1,i}^2 - G_{2,i}^2), \quad
U = \sum_i 2\,G_{1,i}\,G_{2,i}.
\tag{B2}
\end{equation}
The correlation angle is $\theta = \frac{1}{2}\arctan(U/Q)$, and when $p_1$ and $p_2$ are well correlated locally, $\kappa_\rho = \tan\theta = dp_1/dp_2 = d\log n_H / d\log l_c$, which is the density exponent $\kappa_\rho$ by definition.

The identification of $d\log n_H / d\log l_c$ with the three-dimensional radial density exponent $d\log\rho / d\log r$ is calibrated through synthetic observations. Appendix~\ref{app:mdr_aca_validation} applies the same MDR--ACA procedure to structures with known input exponents and recovers their ordering and approximate values.

\section{Validation of the MDR--ACA Density Exponent}
\label{app:mdr_aca_validation}

We test whether MDR combined with ACA can recover spatial changes in the density-profile exponent. The test used a synthetic three-dimensional density field containing a spherical clump and a filament. In this appendix only, we use $\Sigma_m = \mu m_{\rm n} N_{\rm H}$ to denote the mass surface density, to avoid confusion with the column number density $\Sigma$ used in the main text.

\subsection{Synthetic density field}

The total number-density field was written as
\begin{equation}
n_{\mathrm{tot}}(\mathbf{x})
= n_{\mathrm{bg}}
+ n_{\mathrm{ball}}(\mathbf{x})
+ n_{\mathrm{fil}}(\mathbf{x}),
\end{equation}
where $n_{\mathrm{bg}}$ is a uniform background.

The spherical component was defined using a softened power law,
\begin{equation}
n_{\mathrm{ball}}(r)
= n_{0,\mathrm{b}}
\left[\,1 + \left(\frac{r}{r_{0,\mathrm{b}}}\right)^{\!2}\,\right]^{-1},
\label{eq:validation_ball}
\end{equation}
where $r$ is the distance from the centre of the sphere. Outside the central flat region, $n_{\mathrm{ball}} \propto r^{-2}$.

For the filament, $R_\perp$ denotes the perpendicular distance from the filament axis and $s$ denotes the distance along the axis. Its density was defined as
\begin{equation}
n_{\mathrm{fil}}(R_\perp, s)
= n_{0,\mathrm{f}}
\left[\,1 + \left(\frac{R_\perp}{R_{0,\mathrm{f}}}\right)^{\!2}\,\right]^{-1/2}
\left[\,1 + \left(\frac{s}{L_{\mathrm{f}}}\right)^{\!2}\,\right]^{-1}.
\label{eq:validation_filament}
\end{equation}
The first term gives $n_{\mathrm{fil}} \propto R_\perp^{-1}$ at large $R_\perp$, while the second term smoothly reduces the density along the filament axis.

The three-dimensional cube had a pixel size of $\Delta x = 0.02$\,pc. The array axes were ordered as $(z,y,x)$.

\subsection{Projection and MDR reconstruction}

The synthetic column-density map was obtained by integrating the three-dimensional density field along the $z$ direction:
\begin{equation}
N_{\mathrm{H}}(x,y)
= \sum_z n_{\mathrm{tot}}(x,y,z)\,\Delta z.
\label{eq:validation_projection}
\end{equation}
The column density was converted to mass surface density using $\Sigma_m(x,y) = \mu m_{\mathrm{n}} N_{\mathrm{H}}(x,y)$, where $\mu=2.37$ and $m_{\mathrm{n}}$ is the nucleon mass.

MDR was then applied to $\Sigma_m$:
\begin{equation}
\left(\rho_{\mathrm{MDR}}, l_{c,\mathrm{MDR}}\right)
= \mathrm{MDR}\left(\Sigma_m, \Delta x\right).
\end{equation}
The reconstructed density was converted to number density and the characteristic scale to parsecs:
$\langle n_{\mathrm{H}}\rangle = \rho_{\mathrm{MDR}} / (\mu m_{\mathrm{n}})$,
$l_c = l_{c,\mathrm{MDR}} / \mathrm{pc}$.

\subsection{Density exponent from ACA}

ACA was applied to the two MDR maps in logarithmic space, $X = \log_{10}l_c$ and $Y = \log_{10}\langle n_{\mathrm{H}}\rangle$. The local correlation angle $\theta$ was converted to a density exponent using $\kappa_\rho = \tan\theta = \mathrm{d}\log\rho / \mathrm{d}\log r$. Only finite pixels with positive density and characteristic scale were used, with the exponent restricted to $-4 < \kappa_\rho < 0$.

\subsection{Validation result}

\begin{figure*}
    \centering
    \includegraphics[width=\textwidth]
    {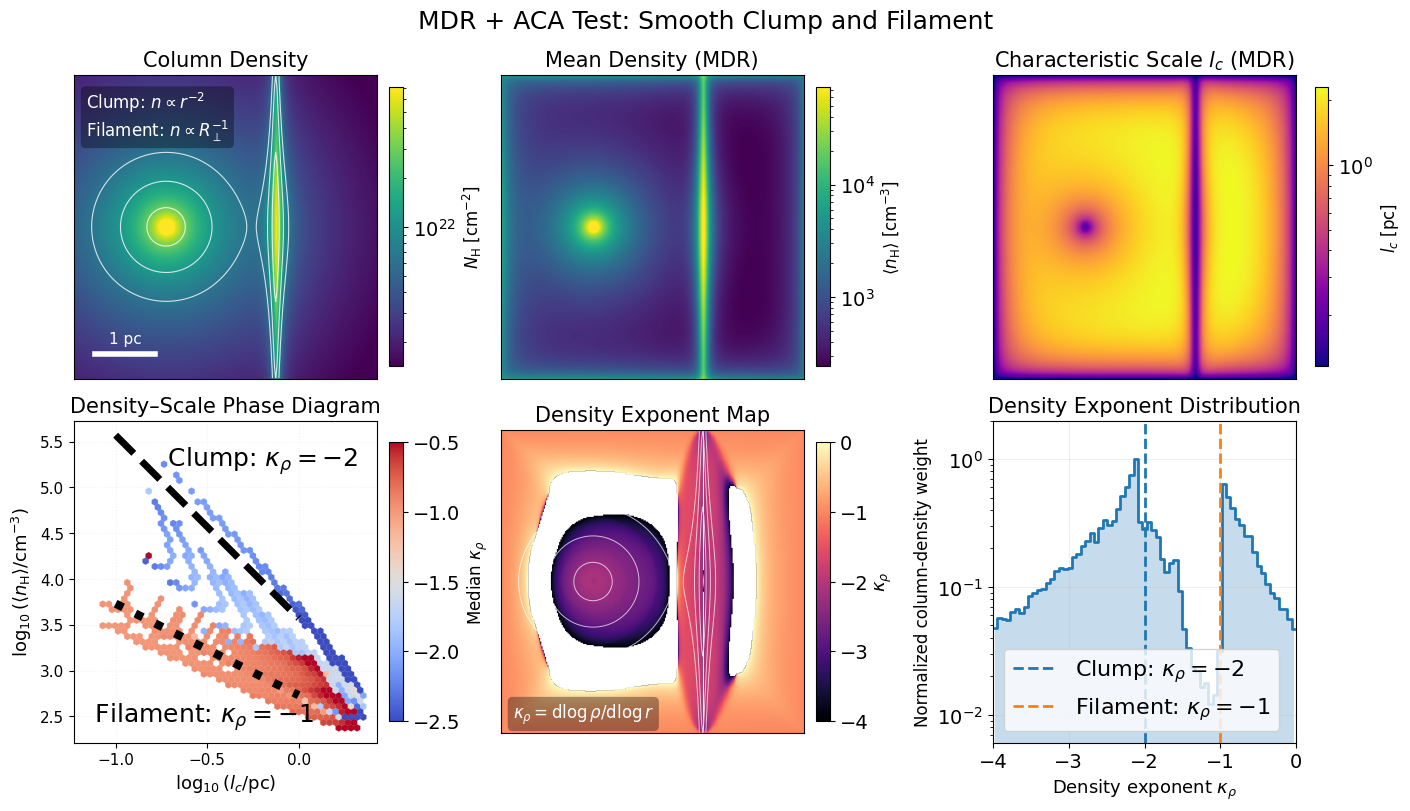}
    \caption{
    Validation of the MDR--ACA density-profile exponent.
    \textit{Top left:} projected column density of the input model ($n_{\mathrm{ball}} \propto r^{-2}$, $n_{\mathrm{fil}} \propto R_\perp^{-1}$).
    \textit{Top middle:} mean number density recovered with MDR.
    \textit{Top right:} MDR characteristic scale $l_c$.
    \textit{Bottom left:} density--scale phase diagram in $\log l_c$--$\log\langle n_{\mathrm{H}}\rangle$ space. Each hexagonal bin is coloured by the median recovered $\kappa_\rho$.
    \textit{Bottom middle:} spatial map of the ACA density exponent.
    \textit{Bottom right:} column-density-weighted distribution of $\kappa_\rho$. The dashed lines mark the input exponents.
    }
    \label{fig:mdr_aca_validation}
\end{figure*}

The two input structures ($\kappa_\rho = -2$ for the sphere and $\kappa_\rho = -1$ for the filament, see Fig.~\ref{fig:mdr_aca_validation}) remain distinct in the MDR-reconstructed density and characteristic-scale maps. Their column-density-weighted $\kappa_\rho$ distribution has two main concentrations near the input values. MDR combined with ACA therefore preserves differences in the input density scaling and recovers the ordering and approximate values of the exponents, consistent with the broader MHD validation reported by \citet{2026ApJ...997..345Z}.

\end{document}